\documentclass[twocolumn,10pt]{IEEEtran}
\usepackage{algorithm,algorithmic,amsbsy,amsmath,amssymb,epsfig,bbm,mathrsfs,multirow,amsthm}
\usepackage{tabu}
 \usepackage{booktabs}
\usepackage{multirow}
 \usepackage[table,xcdraw]{xcolor}
\usepackage{colortbl}
\usepackage{hyperref}
\usepackage{graphicx}
\usepackage[caption=false]{subfig}

\DeclareMathAlphabet{\mathpzc}{OT1}{pzc}{m}{it}

\definecolor{mygray}{gray}{0.6}
\definecolor{myblue}{rgb}{0.8,0.85,1}

\begin{document}

\title{\huge Edge Computing for Semantic Communication Enabled Metaverse: An Incentive Mechanism Design}
\author{
Nguyen Cong Luong, Quoc-Viet Pham,~\IEEEmembership{Member,~IEEE}, Thien Huynh-The, \\ Van-Dinh Nguyen, Derrick Wing Kwan Ng,~\IEEEmembership{Fellow,~IEEE}, and Symeon Chatzinotas,~\IEEEmembership{Fellow,~IEEE} \vspace{-10pt}
}

\maketitle
\begin{abstract}
Semantic communication (SemCom) and edge computing are two disruptive solutions to address emerging requirements of huge data communication, bandwidth efficiency and low latency data processing in Metaverse. However, edge computing resources are often provided by computing service providers and thus it is essential to
design appealingly incentive mechanisms for the provision of limited resources. Deep learning (DL)-based auction has recently proposed as an incentive mechanism that maximizes the revenue while holding important economic properties, i.e., individual rationality and incentive
compatibility. Therefore, in this work, we introduce the design of the DL-based auction for the computing resource allocation in SemCom-enabled Metaverse.  First, we briefly introduce the fundamentals and challenges of Metaverse. Second, we present the preliminaries of SemCom and edge
computing. Third, we review various incentive mechanisms for
edge computing resource trading. Fourth, we present the design of the DL-based auction for edge resource allocation in SemCom-enabled Metaverse. Simulation results demonstrate
that the DL-based auction improves the revenue while nearly satisfying the individual rationality and incentive compatibility constraints.
\end{abstract}


\section{Introduction}
\label{sec:introduction}

Metaverse has recently received extensive interest from both industry and academia~\cite{jeong2022innovative}. By allowing the users to be visually and physically immersed in the virtual worlds from anywhere and anytime, Metaverse has a number of advantages in numerous diversified application scenarios. For example, training autonomous vehicles in the real world faces expensive and risky accidents. Meanwhile, by constructing a digital twin of roads, vehicles, traffic lights and pedestrians, training the vehicles in the Metaverse becomes much safer than that in the real world couterparts~\cite{amini2020learning}. However, Metaverse faces two major challenges. First, to construct accurate digital copies, virtual service providers (VSPs) deploy a massive number of heterogeneous Internet-of-Things (IoT) devices to collect a huge amount of data such that the available bandwidth/capacity is unable to cope with the system communication demands. Second, the VSPs serve a number of users with stringent latency requirements, e.g., $7$ to $15$ ms~\cite{cai2022compute}. This requires fast data processing at the VSPs and low delay in virtual service delivery.

Meanwhile, Semantic communication (SemCom) exploiting deep learning (DL) allows a transmitter to send semantic information, i.e., knowledge, of raw data to its targeted receivers~\cite{xie2021deep}, which helps to reduce the data traffic significantly. Thus, SemCom can well address the problems of skyrocketing demands of data volume and bandwidth in Metaverse. Meanwhile, cloud computing and edge computing are potential solutions that helps the VSPs enabling fast data processing. Compared with cloud computing, edge computing brings the computing resources physically closer to the VSPs~\cite{MaoCST17}. Thus, edge computing can better address the stringent latency requirements of the users, thereby well fitting the Metaverse. As such, SemCom and edge computing are effective solutions to the issues of the Metaverse. However, while SemCom is typically deployed by the VSPs, edge computing resources are provided by edge computing providers (ECPs). In general, the ECP is typically rational, meaning that it has an incentive to provide computing resources if its revenue is guaranteed. Meanwhile, the VSPs will be willing to join the computing resource market if its payoff (utility) is non-negative. Here, the payoff of each VSP is a function of latency caused by the offloading and the cost that the VSP needs to pay the ECP for the computing resources. Therefore, a critical issue is to maximize the revenue of the ECP while guaranteeing
non-negative utilities of the VSPs. Moreover, edge computing capacity is generally limited and thus allocating the limited edge resources to the VSPs is another critical issue. As a result, there is emerging need for the design of incentive mechanisms. 

Auction is an effective approach to gain revenue by assigning resources to buyers who value them most. In conventional auctions, e.g., Vickrey–Clarke–Groves (VCG) auction~\cite{varian2014vcg}, bidders or buyers submit their bids to a seller. Specifically, each bid represents the price that the buyer is willing to pay for a resource. In general, the seller selects buyers with the highest bids as winners and determines prices that the winners pay. However, the conventional auctions are not optimal for maximizing the revenue and guaranteeing economic properties. There are two important properties: individual rationality (IR) and incentive compatibility (IC). Particularly, IR guarantees that the utilities (payoffs) of the bidders are non-negative when joining the auction. Meanwhile, IC guarantees that the bidders receive the highest utilities by submitting their truthful bids. In general, maximizing the revenue and guaranteeing the properties is somewhat conflicting each other and thus is difficult to be solved with traditional algorithms. Fortunately, a recent work~\cite{dutting2019optimal} has shown that DL can well solve complex optimization problems for designing the optimal auction. 

\begin{figure*}[!t]
	\centering\includegraphics[width=0.65\linewidth]{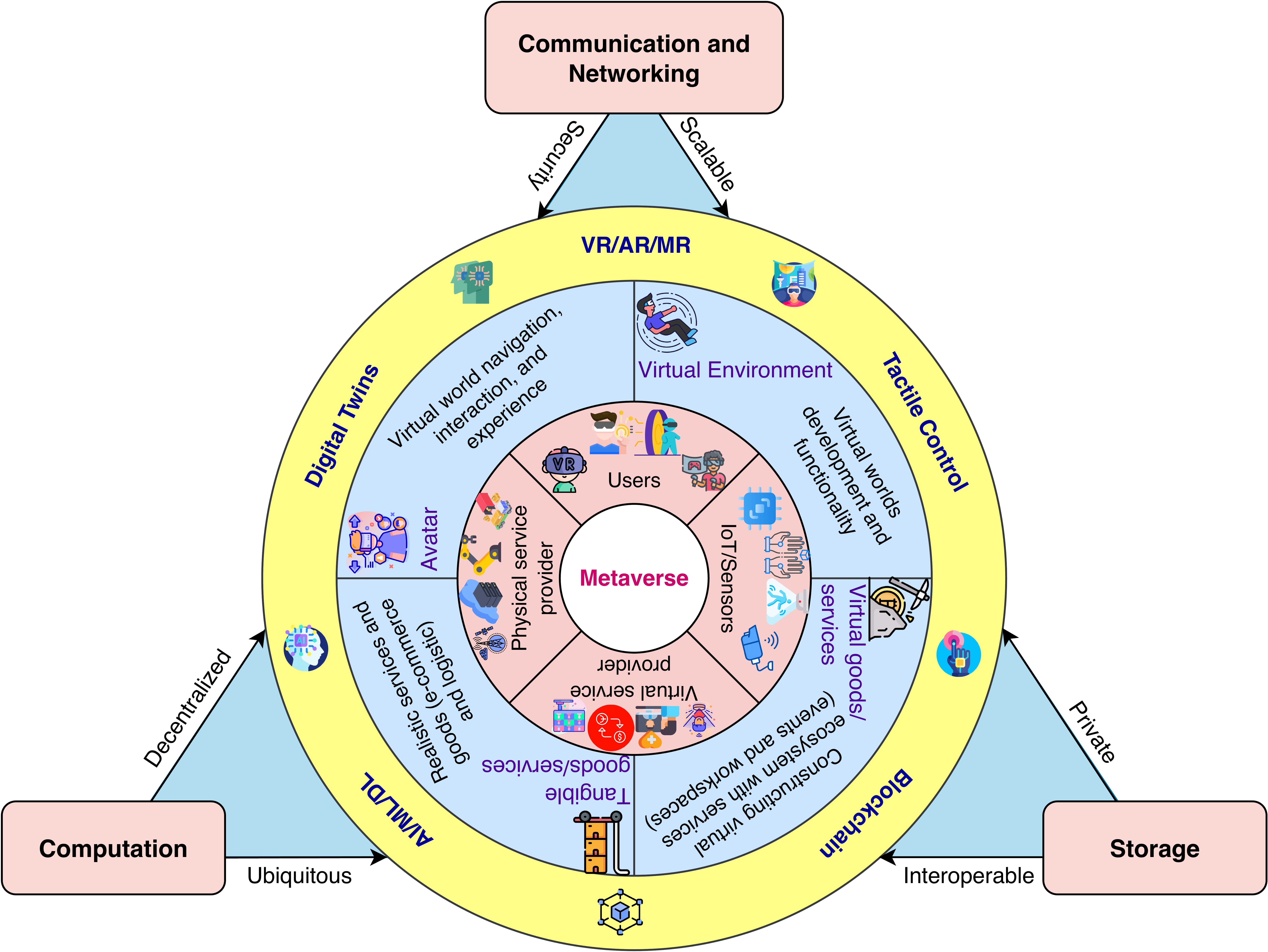}
	\caption{The Metaverse delivers a truly seamless physical-virtual immersive experience driven by various Metaverse engines and different infrastructure sectors.}
	\label{Fig:metaArchitecture}
\end{figure*}

In this paper, we propose a DL-based auction for edge computing resource trading in SemCom-enabled Metaverse systems. First, we briefly introduce fundamentals, applications and challenges of Metaverse. Then, we present the backgrounds of SemCom and edge computing. After that, we review common incentive mechanisms for edge computing resource trading. Next, we consider a case study in which VSPs deploy unmanned aerial vehicles (UAVs) as IoT devices to collect sensing data from physical objects to update their digital twins (DTs). SemCom is adopted to transmit semantic data from the UAVs to their corresponding VSPs. Each VSP can locally compute a part of the data and offload the remaining part fo the ECP. The VSPs compete with each other on the edge computing units deployed by an ECP. An optimal auction for edge computing resource trading between the ECP and VSPs is designed based on DL. The dataset includes valuations of the computing resource units to the VSPs. In particular, the valuations are determined based on the sensing and communication time of the UAVs, the time of data processing at the VSPs and the latency requirements of Metaverse users. The performance evaluation clearly
shows the merit of the DL-based auction and its applicability to edge resource allocation in the Metaverse. Finally, conclusions and interesting research directions are  discussed.

\section{Metaverse: Fundamental and Challenges}
\label{sec:meta}

\subsection{Definitions}

Metaverse, a combination of the prefix ``meta'' and the word ``universe'', is a spatial computing platform that provides immersive digital experiences as an alternative to or a replica of the physical world~\cite{xu2022full}, allowing people to perform real-time interactions with each other and with computer-generated objects, along with numerous diversified economic activities, such as property ownership and trade enabled by blockchain technology.
Relying on a general architecture presented in Fig.~\ref{Fig:metaArchitecture}, the Metaverse is an exemplified version of the Internet that creates an immersive, interoperable, shared, and physical-virtual seamless ecosystem. For ease of presentation, we provide key terms of the Metaverse as follows:
\begin{itemize}
\item Physical-virtual synchronization: Each non-conflicting stakeholder manages some modules that should associate and interact with the virtual world over physical activities. Primary stakeholders are: (1) \textit{Metaverse users} interact with virtual objects, non-player characters, and other users by human-machine interface devices (such as smart glasses, head-mounted display, and gestures); (2) \textit{IoT and UAV-aided sensor networks} collect data from the physical world for synchronizing, maintaining, and modernizing virtual worlds; (3) \textit{Virtual service providers (VSPs)} are responsible for virtual world development and user-oriented service maintenance via built-in Metaverse applications to trigger the applicability and growth of virtual economy; and (4) \textit{Physical service providers} supervise different sectors of the physical infrastructure (including storage, computation, and communication and networking) to support Metaverse engines and deliver tangible goods and services.
\item Metaverse engines gather data and information from stakeholder-controlled modules that associate with the physical and virtual worlds concomitantly for multiple entities administration. There are five principal engines: (1) \textit{Virtual reality (VR)} enables users to experience digital contents in the virtual world invigorated by animation, display, and 3D render technologies; (2) \textit{Tactile control} allows users to interact with other ones and virtual entities in the virtual worlds smoothly and seamlessly through embodied haptic and kinesthetic sensation; (3) \textit{Digital twins (DT)} model multiple-level physical objects by designing digital replicas in the virtual worlds with a full reflection of properties and functionalities; (4) \textit{Artificial intelligence (AI)} plays the roles of automatic analysis and understanding users' behaviors and commands for more accurate assists and recommendations; and (5) \textit{Blockchain} technology with consensus algorithms and smart contracts enables multiple parties to secure virtual goods and assets likes non-fungible token (NFT) through a decentralized economic ecosystem. 
\end{itemize}
The infrastructure layer consisting of communication and networking, computation, and storage enables users to access and immerse the Metaverse at the network edge. For further details on the Metaverse, the interested readers can refer to an extensive review in \cite{xu2022full}.

\subsection{Applications}

By providing seamless physical-virtual interactive and immersive experiences, the Metaverse has a wide range of consumer and industrial applications of various domains.
\begin{itemize}
    \item \textit{Vehicular}: The vehicular Metaverse can be defined as an immersive integration of vehicular communications. The vehicular Metaverse will fuse the virtual environment and the real data (e.g., road, traffic, weather, and motion) to create emerging in-vehicle services and applications for drivers and passengers \cite{jiang2022reliable}.
    \item \textit{Social}: The Metaverse is expected to modernize current online social networks by becoming a meta-hub to connect users from multiple virtual worlds for communication, interaction, and content creation and sharing.
    \item \textit{Gaming}: Powered by VR technology, the entering of the gaming Metaverse has made a massive hype in the gaming industry when the gameplay, environment, and interaction manner should be uplifted to bring ultimate experience to players.
    \item \textit{Healthcare}: The Metaverse has more potential to being leveraged in the medical healthcare and wellness domains with numerous applications, such as telemedicine, collaborative fitness, and remote surgery.
    \item \textit{Industry}: For industrial applications, the Metaverse with built-in DT and IoT can improve the reliability of manufacturing systems, reduce operating risks to quality control and predictive maintenance, and motivate collaboration in process design and product development.
    \item \textit{E-commerce}: The Metaverse will create a great opportunity for consumer brands to provide more delightful shopping experiences and uplift their business models. The borderline between physical shopping experience and those virtual shopping is nearly obliterated in the Metaverse, and behavior understanding-based personalized shopping is the key to boosting the retailers' income.
    \item \textit{Collaboration}: The Metaverse can launch revolutionary working styles and environments, where employees represented as avatars in the virtual worlds can process office tasks, communicate with customers, and collaborate with colleagues conveniently besides other activities such as training and coaching.
\end{itemize}

\subsection{Challenges}
The Metaverse is expected to serve a massive number of users with guaranteed truly immersive user experiences, but it exposes two major issues of unprecedented resource requirements in the infrastructure.

\textit{Communication requirements}: 
In the Metaverse, ensuring ultra-smooth interactions among users and with virtual environment is the first priority, which demands the real-time multiple source data collection, aggregation, and communication in the physical-virtual ecosystem. 
For example, a virtual driving system in the vehicular Metaverse should simulate the virtual road with fresh data collected from the real world. As such, the virtual driving services are realistic and the driving test skills can be continuously practiced. In this regard, we propose deploying IoT devices (e.g., UAVs, smartphones, and ) to sense and collect the real data that will be of interest to the VSP. 
Besides a huge amount of sensory data from the IoT devices in the physical world, the Metaverse has to face with multimedia data from game and video streaming activities in the virtual world, which results in the high bandwidth consumption and high transmission latency, especially in the case of which the Metaverse successively expands applications and services in the future. Consequently, although the end-to-end latency requirements should be stringent, it may still be flexible for different target applications and services to inhibit uncomfortable experiences like lagged responses.

\textit{Computation requirements}: Building, operating, and maintaining virtual worlds with built-in applications and services consume massive computational resources. Besides 3D object modeling and rendering in digital content creation activities, many tasks of image/video processing and computer vision based on DL (for examples, object detection, image segmentation, and scene understanding) require substantial computational complexity. Some Metaverse applications like gaming and healthcare may involve multiple processing tasks (e.g., natural language processing, video rendering, and gameplay recommendation) that usually runs as separated service functions and therefore uses much more computing resources compared with incorporated tasks. Although increasing computational power can reduce computation latency, the required service fee would also increase accordingly and becomes a barrier to reach high-quality services.

To reduce the size of the raw data collected from the physical world, SemCom has emerged as an effective solution, and to reduce the data processing time and guarantee the latency requirements of the Metaverse users, edge computing is adopted.

\section{Semantic Communication and Edge Computing for Metaverse}
\label{sec:seman_edge}

\subsection{Semantic Communication (SemCom)}
\label{sec:seman}

\begin{figure*}[t]
	\centering
	\includegraphics[width=0.95\linewidth]{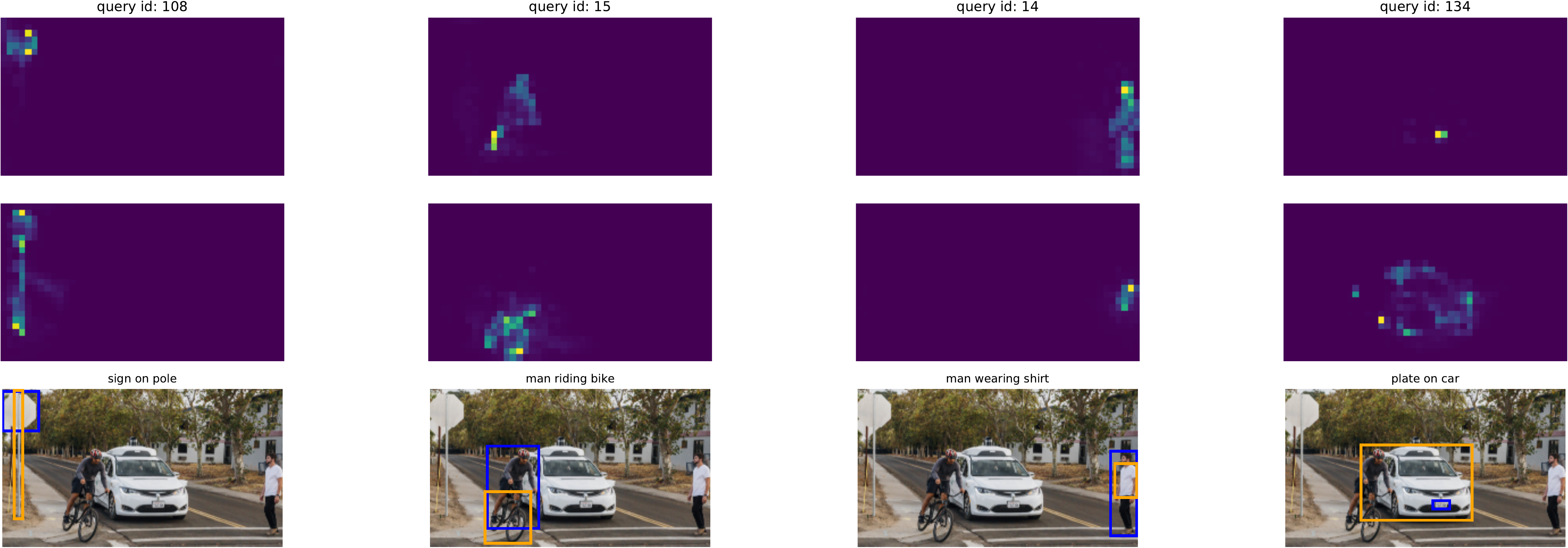}
	\caption{Illustrative example of the use of SemCom for the Metaverse.}
	\label{Fig:semcom_result}
\end{figure*}

SemCom is a promising technology for future IoT and wireless systems. In contrast to conventional communication systems that rely on the Shannon information theory, SemCom only transmits relevant information to the targeted applications adopted at the receiver. Instead of sending bit sequences representing the whole raw data, SemCom allows end devices to extract the features relevant to the Metaverse tasks. As such, irrelevant information about the source data can be omitted and is not transferred to the server without sacrificing the performance. Notably, conventional communication systems only focus on the technical level in the three-level communication paradigm proposed by Weaver \cite{weaver1953recent}. SemCom is opposed by also integrating the semantic and effectiveness levels into the design, thus making SemCom more efficient for emerging services and applications. 


The fact is that Metaverse services and applications generally create massive data, raising the issue of efficient data processing and placing a big burden on the data-centric paradigm in today's networks. Therefore, SemCom has the potential to serve a key enabler in Metaverse. In particular, the data collected by end devices, such as UAVs, head-mounted displays, and AR devices, can be processed first to extract the semantics. For example, unlike conventional UAV-assisted sensing systems that require the UAVs to transmit all the captured images to the edge server for learning the tasks of Metaverse services, the images collected by UAV-assisted sensing Metaverse systems are fed into the attention-based semantic encoder to choose selectively the interested images for ground users with individual preferences \cite{kang2022personalized}. This allows the UAV only to transmit highly interesting images to the users so as to increase the match score, reduce the latency, and save bandwidth resources. Moreover, efficient transmission of semantic data in multi-user networks is hindered by wireless channel dynamics and limited network resources (e.g., UAV energy budget, bandwidth, and computing resources). Accordingly, it is necessary to design efficient resource allocation in SemCom-enabled Metaverse.

Fig.~\ref{Fig:semcom_result} presents our preliminary results of semantic extraction from a real dataset~~\cite{UAV_dataset} related to vehicular Metaverse. The images of the dataset can be collected by smart vehicles or UAVs empowered by embedded sensors and cameras. The original image with a size of $3.59$ Mbytes is fed into a SemCom system. The outputs of the SemCom system include the boxes and their corresponding SemCom texts, e.g., ``sign on pole", ``man riding bike", and ``man wearing shirt". The total size of the boxes is only $0.65$ Mbytes and that of the texts is $56$ bytes, which is much lower than the size of the original image, i.e., $3.59$ Mbytes. These results demonstrate that the use of SemCom can significantly reduce the size of the raw data. The outputs of the SemCom are then transmitted to the VSP for further processing. Given the small data size, the VSP can locally process the data or offload the data to an edge computing platform without the need for remote cloud computing. 
\subsection{Edge Computing}
\label{sec:comp_off}

Cloud computing provides computing services (e.g. storage, databases, servers, software and analytics) to end users via virtual machines. To cope with an unprecedented generation of huge amounts and heterogeneous modalities of data, fog computing is proposed to decentralize the cloud infrastructure to  provide services to the users on edge networks, thus reducing the \textit{round-trip} latency and improving connectivity \cite{ChiangIoTFog16}. On the other hand,  edge computing aims to bring computing and data storage closer the physical network's edge of either IoT devices or the data source where it is being generated and collected \cite{MaoCST17}, providing real-time and high computation-intense capabilities, reducing bandwidth needs as well as preserving data privacy.

The Metaverse will digitally create a new virtual world to replicate all aspects of the physical environment. To make the Metaverse a reality, advanced and computation-intense capabilities of computing severs are required to deliver near ``zero'' latency services \cite{ChengNET22}. This may no longer be feasible by relying solely on the traditional cloud computing that can be possibly thousands of miles away. Otherwise, as presented in Section~\ref{sec:seman}, SemCom successfully transmits and/or exchanges semantic information relevant to the specific tasks at the receivers rather than the transmission of raw data, which significantly reduces the collected data at the VSP. As a result, the VSP may not need multi-layer computing  approaches that would require an enormous amount of additional complexity on top of existing power-hungry and over-sized communication networks. Therefore, we envision a new paradigm by enabling edge computing for SemCom-based Metaverse. This paradigm allows the Metaverse platform to directly process tasks at the network edge, thus reducing bandwidth demands, minimizing network latency  and storing significant data locally.

However, edge computing service providers (ECPs) are typically individual and rational, meaning that they have an incentive to provide computing resources only when their revenue is guaranteed. Thus, it is essential to design incentive mechanisms that encourage ECPs to provide edge computing services, making Metaverse to be responsive and lifelike.


\subsection{Incentive Mechanisms for Edge Computing}
\label{sec:incen_edge}
Game theory and auction are two common incentive mechanisms. Given the edge resource constraint, auction is an effective solution since it allocates resources to the buyers who value them the most. The following present common auctions adopted for encouraging edge computing trading.
\subsubsection{First-price auction and second-price auction} In both the auctions, the VSPs acting as bidders submit their bids to the ECP to compete on an edge computing resource. The bid is the price that the VSP is willing to buy the resource. The VSP with the highest bid is the winner. Then, with the first-price auction, the winning VSP pays the ECP the bid that it submits, while with the second-price auction, the winning VSP pays the ECP with the second highest bid. Thus, the first-price auction allows the ECP to receive higher revenue than the second-price auction. However, the second-price auction guarantees certain desired properties, i.e., IC and IR, since the VSP pays the seller a price lower than that it submits.

\subsubsection{VCG auction}
VCG auction~\cite{varian2014vcg} allows each VSP to submit its bids for multiple edge computing resources. The VSPs with the highest bids are the winners of the resources. Then, for each winning resource, the VSP pays the ECP a price that is the difference between the social welfare value if the VSP does not participate in the auction and the attainable social welfare after the VSP wins the resource. By this way, the VCG auction is proved as a generalization of the second-price auction, which also guarantees the IC and IR properties.

\subsubsection{Revenue-optimal auction}
The VCG auction guarantees IR that motivates the VSPs to participate in the auction. However, by charging the winning VSPs in an optimally social welfare, the revenue obtained by the ECP is low or even zero. Thus, the ECP may have no incentive to contribute resources. There is always a trade-off between the revenue maximization and the IR guarantee. Thus, designing an optimal auction in terms of maximizing revenue, subject to IR and IC is necessary. However, this problem is a non-convex problem, which is very challenging to be solved. Fortunately, deep learning (DL) using stochastic gradient descent has been recently shown to effectively solve optimization problems, and it is particularly used to design the nearly-optimal auction~\cite{dutting2019optimal}. Note that the revenue-optimal auction in~\cite{dutting2019optimal} is proposed for general scenarios.


\section{A Case Study: Edge Computing Trading for SemCom Enabled Metaverse}
\label{sec:use_case}
In this section, we introduce an edge computing trading for a SemCom-enabled Metaverse system. Then, we present the use of the DL-based auction~\cite{dutting2019optimal} for trading. Numerical results are provided to show the effectiveness of the DL-based auction.

\subsection{System Model}
\label{sec:system}
\begin{figure}[h!]
	\centering
	\includegraphics[width=0.95\linewidth]{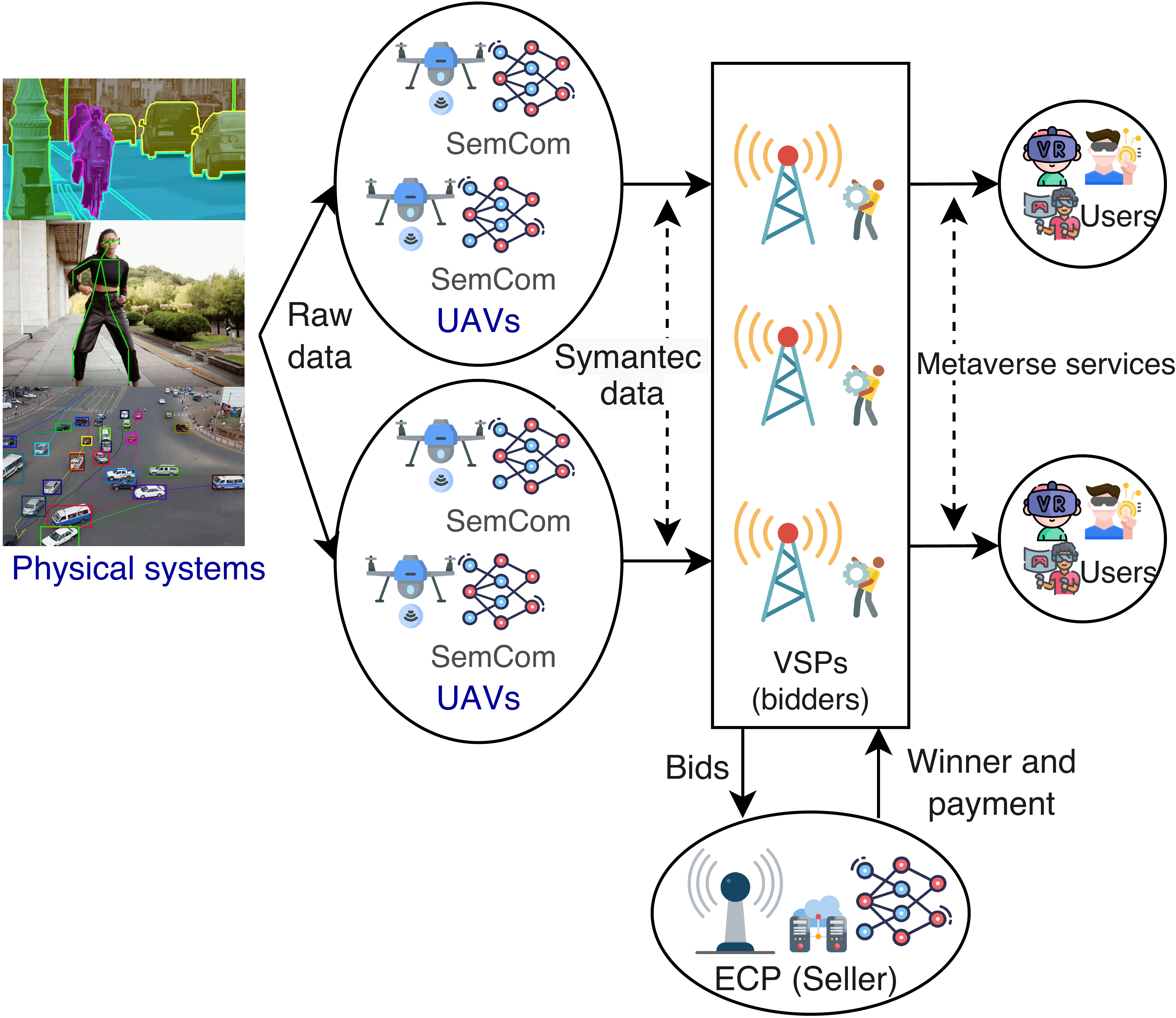}
	\caption{Edge computing trading for SemCom-enabled Metaverse system.}
	\label{UAV_VSP_AUCTION}
\end{figure}
Figure~\ref{UAV_VSP_AUCTION} shows an edge computing-enabled Metaverse system that consists of $N$ VSPs. The Metaverse system may be a virtual driver
training system. Each VSP $n\in\{1,\ldots, N\}$ serves its $A_n$ Metaverse users. For this, it deploys $M_n$ UAVs for sensing a physical system, e.g., car traffic condition, to collect its features. The UAV performs the sensing for $t_{n,m}^{\rm{Sens}}$ seconds, and then they use the SemCom technique to extract and wirelessly transmit the semantic data to its corresponding VSP. The UAV needs $t^{\rm{Comm}}_{n,m}$ seconds to transmit the data to the VSP. In general, $t^{\rm{Comm}}_{n,m}$ is proportional to the size of the semantic data, allocated bandwidth, distance, and quality of the fading channel between the UAV and VSP. The VSP waits to receive the data from all its UAVs before further processing. The reason is that the DT, e.g., the virtual driver
training system, is created from a combination of features from physical objects such as roads, traffic, and weather of the physical system. For this, VSP $n$ requires $t^{\rm{Sens-Comm}}_{n}$ seconds, which is actually the time that the VSP receives from the slowest UAV. 

After that, the VSPs perform data processing that includes data cleaning, classification, and modeling. To process the collected data, each VSP spends $t^{\rm{Comp}}_n$ seconds, which is proportional to the total number of bits received from its UAVs, its available CPU resource $f_n$, and complexity of data processing $\zeta$. Therefore, the computation latency at the VSP is $t^{\rm{Sens-Comm}}_{n} + t^{\rm{Comp}}_n$. In pratice, the $A_n$ Metaverse applications have different latency requirements, and if the application with the lowest latency requirement, denoted by  $t^{\rm{Req}}_{n}$, is satisfied, the remaining applications should be satisfied. Therefore, the total latency caused by the data sensing, communication, and processing must be less than $t^{\rm{Req}}_{n}$, i.e., $t^{\rm{Tot}}_n \leq t^{\rm{Req}}_{n}$. Due to the heterogeneity of Metaverse applications, the condition is not always satisfied. Thus, each VSP can process a part of data collected from its UAVs and requests a computing offloading service an ECP to process the remaining part.
\subsection{Computing Service Trading Market}
\label{sec:compu_trading}
The ECP offers $M$ computing units, e.g., virtual machines, as the items to the VSPs. The VSPs as bidders, i.e., buyers, compete on the computing units by submitting their bids to the ECP. Each bid is a price that the VSP is willing to pay the ECP for a computing unit. Note that the VSP can submit and win more than one computing unit to facilitate a faster process on its data. Based on the bids, the ECP formulates an optimization problem that determines the winning VSPs of computing units and the prices that they need to pay the ECP. The objective is to maximize the revenue while guaranteeing both the IR and IC properties. It is challenging to solve the revenue optimization
via conventional optimization methods due to the generally non-convex objective and constraints. Moreover, the bids of the VSPs as the input parameters of the problem vary over time due to the dynamics of the UAV environments and the latency requirements of the Metaverse. As such, DL using
stochastic gradient descent can successfully find
globally optimal solutions for the constrained optimization problems~\cite{dutting2019optimal}. Moreover, the trained DL can decide online the winning VSPs and their payments. Thus, we resort to the DL approach~\cite{dutting2019optimal} to adaptively search for the solution such that
the convergence to and the optimality of the obtained solution
can be guaranteed. 
\subsection{Learning Auction}
\label{sec:opt_auc}
We define the following terms: 
\begin{itemize}
\item \textit{Valuation:} This represents how much a VSP is willing to buy a computing unit. Denote $v_n$ as the valuation of VSP $n$. Then, if $t^{\rm{Tot}}_n \leq  t^{\rm{Req}}_{n}$, the latency requirements of all the applications are satisfied. In this case, the VSP does not buy any computing unit and $v_n=0$. If $t^{\rm{Tot}}_n > t^{\rm{Req}}_{n}$, the VSP is more willing to buy the computing units and thus we can express $v_n$ as $ t^{\rm{Req}}_{n}- t^{\rm{Tot}}_n$. 
\item \textit{Utility:} Utility of each VSP is the difference between the multiplication of probabilities of winning computing units and the valuations of the VSP and the total price that the VSP pays the ECP for winning the computing units. 
\item \textit{Expected revenue:} This is the total price that the winning VSPs pay the ECP.
\item \textit{IR penalty:} This happens if the utility of any VSP is negative. The IR penalty is thus expected to be zero.
\item \textit{IC penalty:} This is the maximum gain in utility achieved by any VSP if it submits an untruthful bid, i.e., the bid is different from the valuation. The IC penalty is expected to be zero to guarantee the IC property.
\end{itemize}

The expected revenue is the objective, and IR and IC penalties to the VSPs are the constraints. The learning auction is implemented by using two feed-forward neural networks (FNNs). The output of the first FNN includes the winning probabilities of the VSPs, and that of the second FNN includes the prices charged to the VSPs. A common loss function for both the FNNs is formulated based on the augmented
Lagrangian method that allows the constraints
to be enforced through introducing some weighted terms to
the objective~\cite{dutting2019optimal}. Specifically, the common loss function is proportional to the IR and IC pelanties and inversely
proportional to the expected revenue. The valuations of the VSPs as dataset, and the two FNNs use the valuations as their inputs. Note that the valuations of the VSPs are private and unknown to the ECP, but it can obtain the valuations by its historical observation. After that, the two FNNs are trained simultaneously by updating the FNNs' weights and the weights associated with the constraints to minimize the loss function. The loss function depends only on the input without the reference value, the learning algorithm is unsupervised learning.

\subsection{Numerical Results}
\label{sec:num_result}
We provide simulation results to evaluate the effectiveness of the proposed DL-based auction scheme. The number of VSPs is $5$ and each VSP deploys $2$ UAVs that collect the sensing data from the streets. The sensing time is assumed to be $2$ seconds, and the sensing rate of each of the UAVs is $3$ images/second. For SemCom, we leverage a real dataset, namely UAV123~\cite{UAV_dataset}, including images taken from low-altitude UAVs. Then, the images are randomly taken and fed to a pre-defined deep learning model, which is the RelTR online demo~\cite{cong2022reltr}, to extract the corresponding semantic symbols. The extraction time is ignored due to the real-time processing of the pre-defined deep learning model. The semantic symbols are then converted into bits and wirelessly transmitted to the corresponding VSPs. The available CPU resources at the VSPs are within $[5,10]$ GHz. The data processing complexity is randomly distributed in the range of $600$ cycles/bit. The number of Metaverse applications that each VSP serves is $3$. The latency requirements of Metaverse
applications are within $[1,3]$ seconds. For comparison purpose, we adopt the well-known VCG auction~\cite{varian2014vcg} as a baseline scheme.


\begin{figure}[!ht]
	\centering
	\subfloat[\label{Fig:Revenue}]{\includegraphics[width=0.7\linewidth]{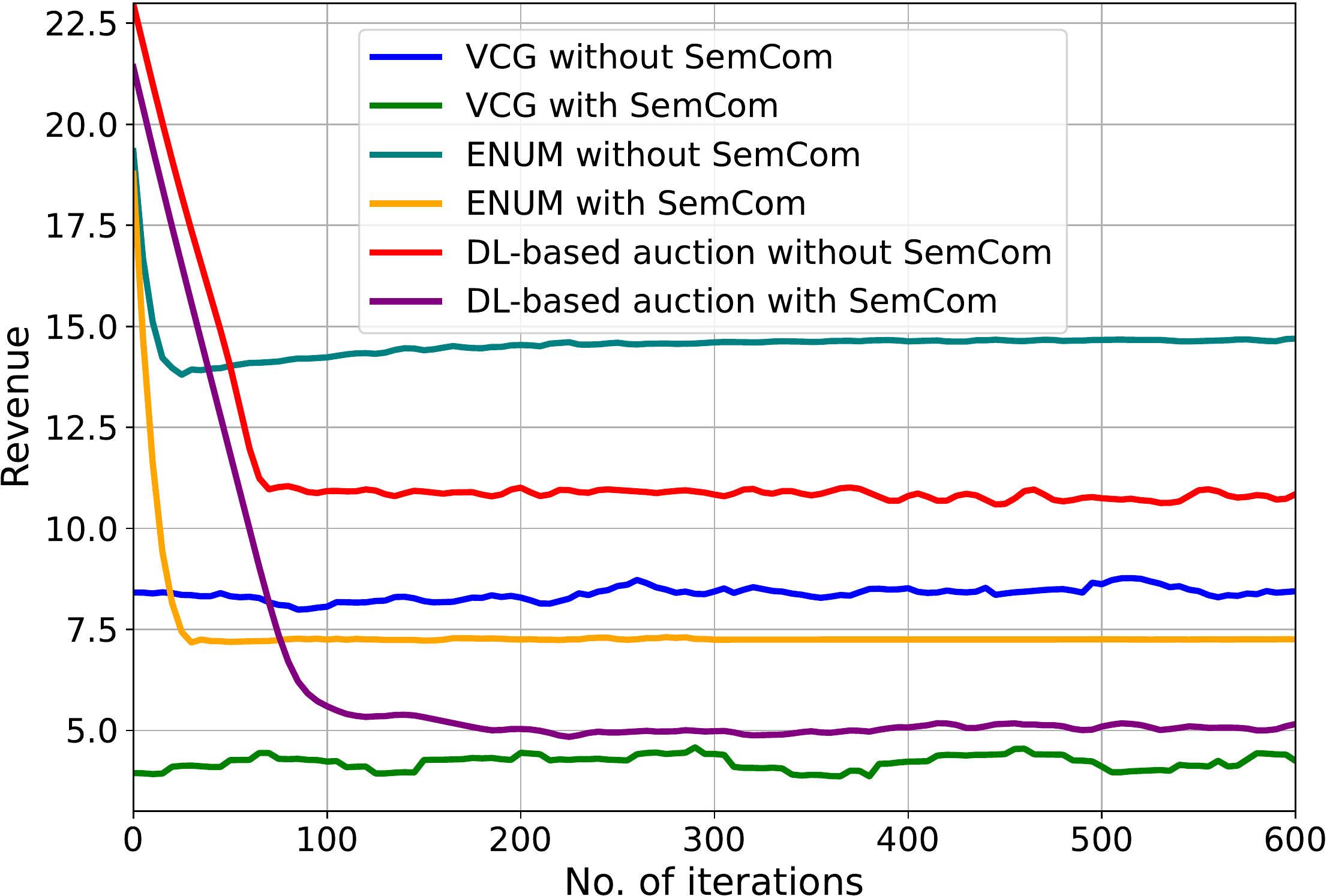}}\qquad
	\subfloat[\label{Fig:IR_IC}]{\includegraphics[width=0.7\linewidth]{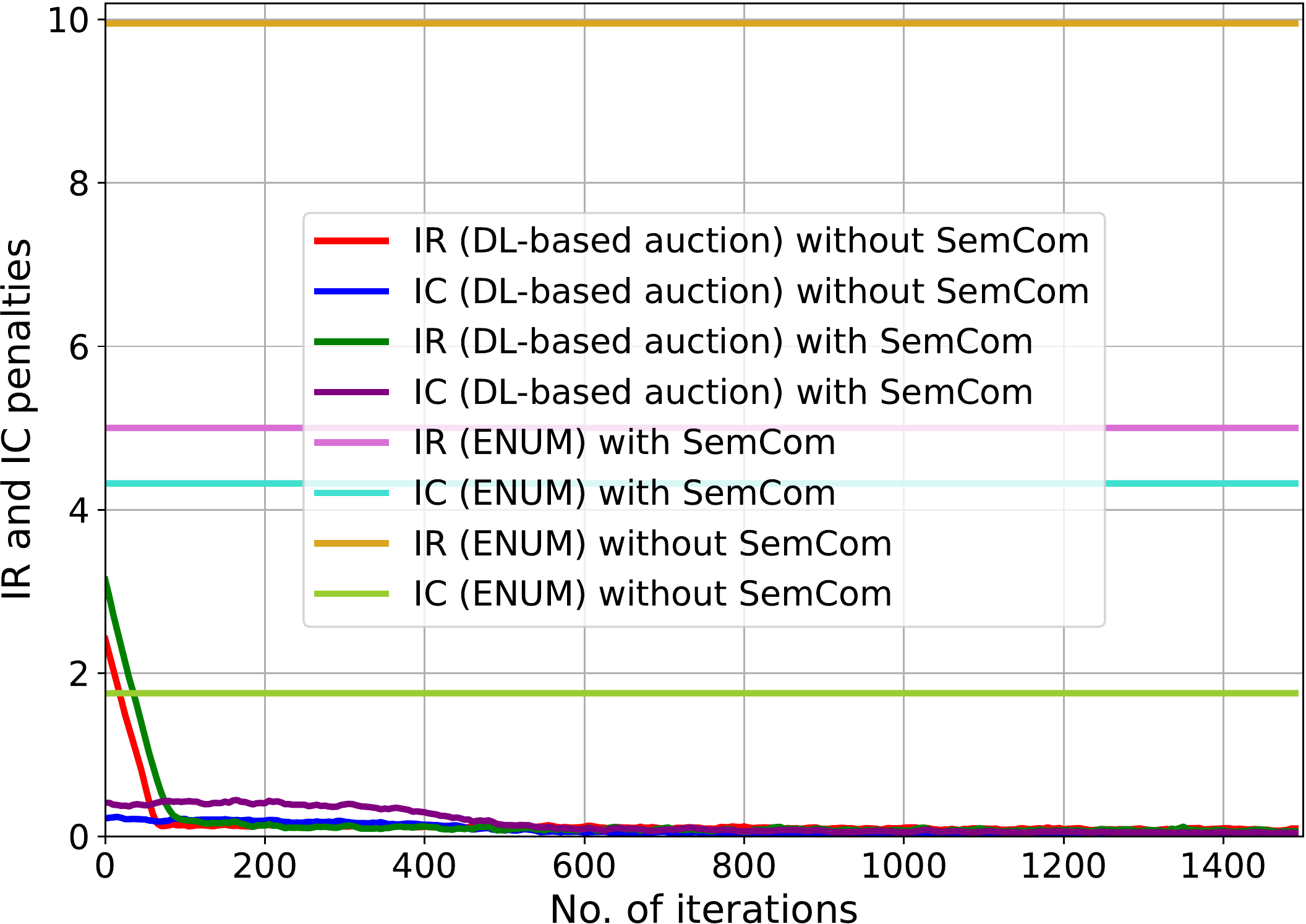}}
	\caption{(a) Convergence of the DL-based auction and (b) IR and IC penalties.}
	\label{Revenue_IR_IC}
\end{figure} 

First, we discuss the convergence of the algorithms. As shown in Fig.~\ref{Revenue_IR_IC}(a), the DL-based auction  with and without SemCom are able to converge to their stable values. Moreover, the revenue obtained by the DL-based auction is much higher than that obtained by the VCG scheme. The reason is that the VCG scheme aims to maximize the social welfare rather than the revenue. Importantly, compared with the case that the VSPs do not use SemCom, the revenue achieved by the ECP with SemCom is much lower. The reason is that as SemCom is used, the semantic symbols with small sizes (rather than the raw images) are transmitted to the VSPs. This results in decreasing the valuations of the VSPs, i..e, the VSPs are less willing to buy the computing units, and the ECP's revenue. Note that the revenue is the total price paid by the VSPs. As such, the use of SemCom significantly decreases the payment of the VSPs for the computing units. 

Next, we show how the proposed DL-based auction guarantees the IR and IC. As shown in Fig.~\ref{Revenue_IR_IC}(b), in both the cases with and without SemCom, the IR and IC penalties achieved by the DL-based auction are almost zero. Recall that as the IR penalty is zero, the utility of the VSPs is nonnegative as participating in the auction. Meanwhile, the zero IC penalty means that the VSPs have no incentive to submit their untruthful bids.

\begin{figure}[h]
	\centering
	\includegraphics[width=0.7\linewidth]{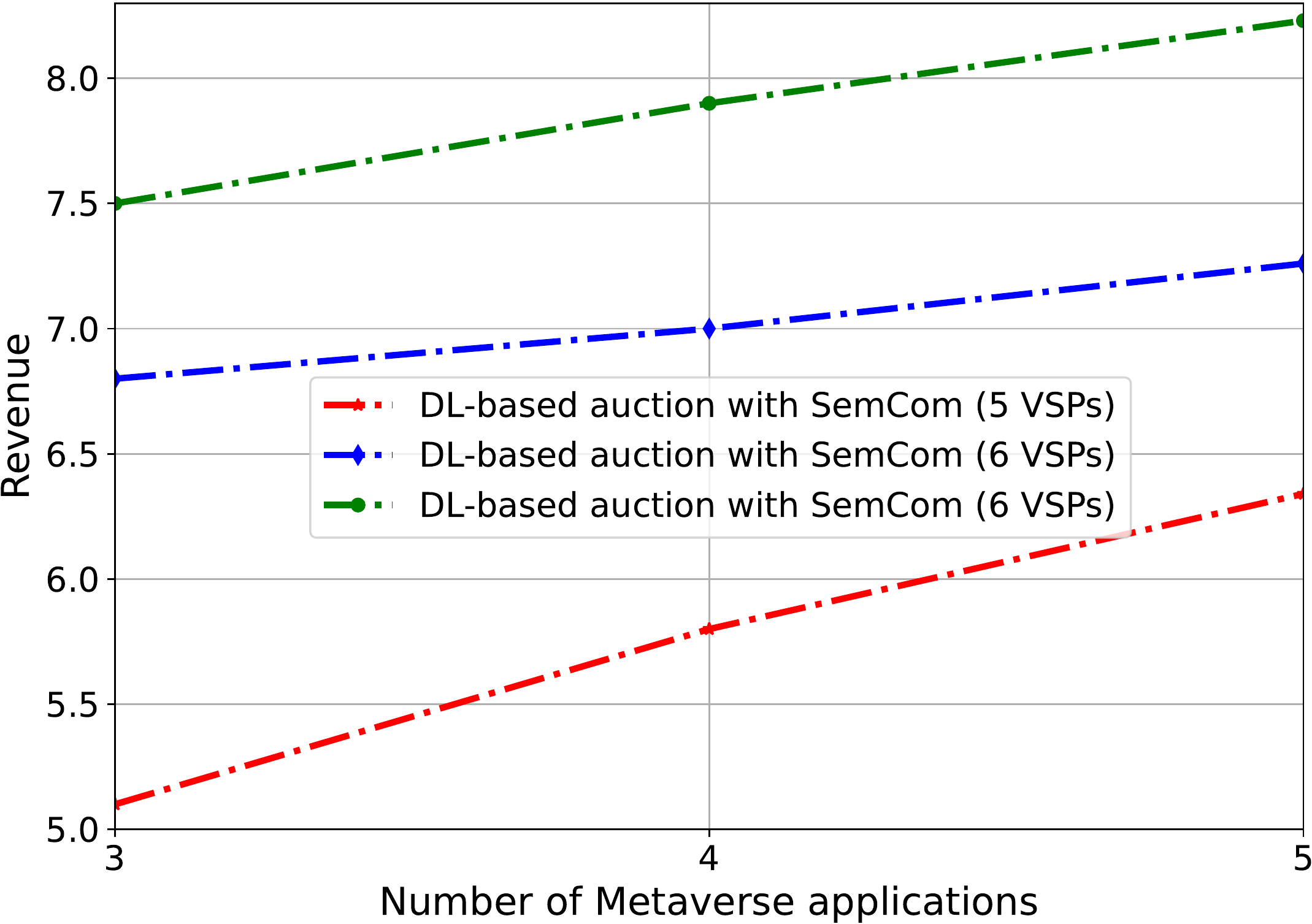}
	\caption{Revenue versus the number of Metaverse applications and VSPs with SemCom.}
	\label{An_VS}
\end{figure}

Finally, it is worth discussing the impacts of the number of VSPs and Metaverse applications on the ECP's revenue. As shown in Fig.~\ref{An_VS}, the revenue increases as the number of VSPs increases. The reason is that the high number of VSPs results in high competition among them, and thus only VSPs with high valuations are the winners. The revenue also increases as the number of Metaverse applications increases. This is because there is a high probability that the VSPs will serve the Metaverse applications with low latency requirements. To satisfy such an application, the VSPs are more willing to buy the computing units, resulting in high valuations. 

\section{Conclusions and Future Research Directions}
\label{sec:conc.}

In this paper, we propose a DL-based auction for edge computing trading in a SemCom-enabled Metaverse system. The proposed auction aims to maximize the revenue of the ECP and achieve the desired economic properties, i.e., IR and IC. This is to motivate both the ECP and VSPs to provide their services to the market. The simulation results clearly show that the proposed auction significantly improves the revenue compared with the baseline while achieving nearly zero IR and IC penalties. Based on our proposed system model, future research directions can be investigated as follows:  
\begin{itemize}
\item To fully serve the VSPs, multiple ECPs can be deployed. The ECPs are different in computing resources, network resources, and incentive costs which leads to different payoffs for the VSPs. The major problem is how to assign an ECP and a VSP to maximize the utilities of both ECPs and VSPs. Matching theory or double auction can be exploited to effectively solve the problem.
\item Each VSP can optimize the data size for offloading to minimize the total latency and incentive cost. However, the available computing and network resources at edge computing are dynamic and uncertain. In particular, at each current time step, each VSP may not know the resources in the next time steps depending on the offloading decisions of other VSPs. In this case, multi-agent learning algorithms can be used as an effective tool to solve the problem.
\item UAVs may not be owned by the VSPs and they can be deployed by a third party, e.g., a sensing data provider. In such a case, the sensing data market between the VSPs and the sensing data provider needs to be investigated, in addition to the computing market between the VSPs and the ECPs. To solve the multi-level interaction, hierarchical auctions or games are effective tools.
\end{itemize}

\bibliographystyle{IEEEtran}
\bibliography{Ref}

\end{document}